\documentclass[journal]{IEEEtran}

\usepackage{cite}

\usepackage{graphicx}

\usepackage{amsmath}
\usepackage{amssymb}
\usepackage{subfigure}

\newtheorem{theorem}{{Theorem}}
\newtheorem{definition}{{Definition}}
\newtheorem{remark}{{Remark}}
\newtheorem{lemma}{{Lemma}}
\newtheorem{corollary}{{Corollary}}
\newtheorem{assumption}{{Assumption}}

\usepackage[overload]{empheq} 
\usepackage{algorithm,algorithmic}
\usepackage{xcolor}

\usepackage{subfig}

\usepackage{tikz}
\usepackage{tkz-graph}
\usepackage{circuitikz}
\usetikzlibrary{graphs,,positioning,fit,backgrounds}
\usetikzlibrary{arrows.meta}
\usetikzlibrary{babel}
\usetikzlibrary{decorations.markings}
\tikzset{>={Stealth[length=2mm]}}
\tikzset{->-/.style={decoration={
  markings,
  mark=at position #1 with {\arrow{>}}},postaction={decorate}}}

\usepackage{url}

\begin{document}
%


\title{Simultaneous Detectability of Process and Sensor Faults: Application to Water Distribution Networks}

%
%
%

\author{Krishnan~Srinivasarengan, 
        Taha~Boukhobza,
        Samir~Aberkane 
		and~Vincent~Laurain
\thanks{The authors are with the Université de Lorraine, CNRS, CRAN, F-54500, Nancy, France. Corresponding author: vincent.laurain@univ-lorraine.fr}
\thanks{The research work was supported by the SPHEREAU project}
}

\markboth{IEEE Transactions on Control Systems Technology}%
{Shell \MakeLowercase{\textit{et al.}}: Bare Demo of IEEEtran.cls for IEEE Journals}
%


\maketitle

\begin{abstract}
	Detecting leaks in Water Distribution Networks (WDN) using sensors has become crucial towards an efficient management of water resources. The leak detection methods that use this data rely on the correctness of the acquired data. However, this assumption is often violated in practice. Consequently, leak detection under sensor faults is a problem of practical importance. This relates to the more general problem of simultaneous detectability in sensor and process faults for a class of systems modelled as a network, by exploiting the redundancies available through the topological relationship between the sensors. This paper hence aims at i) modeling WDN as graphs containing both systems and sensors faults ii) providing theoretical joint detectability results for such graphs and iii) applying these results to the scenario of leak identification under sensor faults conditions on real data issued from a rural WDN.
\end{abstract}


%
\IEEEpeerreviewmaketitle

\section{Introduction and Motivation}

Water distribution networks (WDN) are an integral part of the urban and rural infrastructure. Water loss due to leaks is one of the significant contributors to the inefficient operation of a WDN. For example, France suffers from distribution losses of nearly $20\%$ and with more than $1000m^3$ of loss per kilometre of pipelines every year \cite{eureau2017}. Hence the problem of leak detection has long been of interest to researchers and engineers.


The interest in remote detection of leaks in water distribution network grew to prominence with the installation of sensors in the WDN. These techniques evolve along with the theoretical developments in the control-theoretic and data analysis community. A standard approach to leak detection in the literature deploys flow sensors at the entrance of DMAs (District Metering Area) and a host of pressure sensors at the pipe junctions. The former helps in detecting the faults while the latter helps to localize them. Model-based approaches have been quite popular in the literature (see for instance, \cite{perez_leak_2014}, \cite{vento_leak_2009}, \cite{vrachimis_leak_2018}). The premise largely rests on developing a model of the network and then develop appropriate observers or estimation techniques that can detect and isolate leaks. The developed solution is then typically tested on an EPANET \cite{rossman1999epanet} model of a real water network or on data from a real network (e.g., \cite{perez_leak_2014}). Data-based models feature prominently in the literature in recent times. This includes the use of Support Vector Machine (SVM) models in\cite{eliades_leakage_2012,brentan2018infer}, Artificial Neural Network (ANN) models in \cite{romano_near_2012}), a graph-based spectral clustering in \cite{candelieri_graph_2014}, and an innovative leak signature space approach for leak localisation in \cite{casillas_leak_2015}. \cite{rajeswaran_graph_2018} proposes an intuitive approach with a multi-stage graph partitioning aids in progressively localisation faults.

When analysing a WDN during its normal operation, one faces sensor faults that manifest in various forms, such as missing data, measurement drift/bias, stuck at a value, etc. Unfortunately, the leak detection literature does not typically consider sensor faults, since simultaneous system/sensor fault detection requires specific assumptions, whether they are temporal conditions \cite{dunia1998joint}, physical modeling assumptions \cite{keliris2015distributed}  or probabilistic assumptions \cite{krishnamoorthy2015simultaneous}.

It is evident from the brief literature survey that an interconnected nature of the underlying system offers an opportunity to distinguish the sensor and process faults. This takes importance in the context of the rise of large scale and cyber-physical systems \cite{allgower2019position}, where a network of sensors monitor the performance of the system. In the literature, one can find the use of graph-theoretic models for various problems in WDNs such as the analysis of structural robustness \cite{yazdani2012applying}, observability analysis \cite{diaz2017topological}, leak detection \cite{carpentier_state_1991}, model decomposition \cite{deuerlein2008decomposition}, and sensor placement \cite{shiddiqi2017sensor}. Our interest lies in understanding the structural properties of a system such as the WDN, as treated in pioneering works such as \cite{carpentier_state_1991} and more recent work as \cite{diaz2017topological}.

\begin{figure}
\centering
\includegraphics[trim={0cm 0.2cm 5.5cm 21.5cm},clip,width=0.9\linewidth]{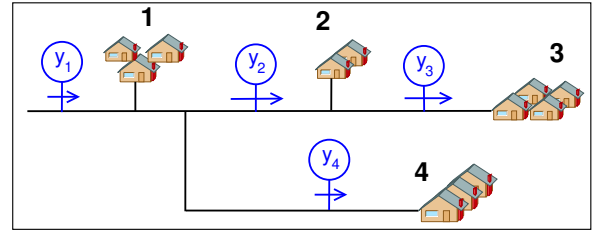}
\caption{A schematic of a part of a WDN}
\label{fig_tree3_schem}
\end{figure}

This paper exploits the natural interconnected structure of the WDN as in Fig~\ref{fig_tree3_schem}, and the redundancy in sensors that arises as a result. In this example, each $y_i$ represents a flow sensor, and it is clear that any water flowing through $y_3$ must flow through $y_2$ and $y_1$ beforehand. This realistic physical representation naturally leads to the well-suited tools of graph.

The aim of this paper can hence be summed up as: is it possible, using a sensor network with certain redundancies, to distinguish system faults from sensor faults using graph theory theoretic tools? The results presented in this paper has its origins on the analysis on a rural WDN instrumented with flow sensors, and where the process faults represent the leaks or abnormal consumption. Nevertheless, the problem of simultaneously distinguishing process and sensor faults in an interconnected system is more generic, and can encompass a wide range of problems. 
 
The main contributions of the paper can be summarized as:
\begin{itemize}
\item Novel use of topological methods to capture the process and sensor faults through graph models 
\item Graph-theoretic conditions for the (simultaneous) detectability of the process and sensor faults
\item As an application of the detectability conditions, an algorithm to  estimate of leaks in a rural WDN
\end{itemize}
In Sec.~\ref{sec_Prelim}, we introduce some preliminary notations and definitions, in particular the graph theoretic concepts used in this paper. Sec.~\ref{sec_graph_modeling} provides a graph modelling of WDN. In Sec.~\ref{sec_problem_formulation}, the detectability problem is presented along with the derivation of the graph-theoretic detectability conditions, which is the core contribution of the paper. Based on the presented results, a novel algorithm for an explainable leak estimation is provided in the Sec.~\ref{sec_leak_detection} and is tested both on a simulated dataset and real dataset obtained from a section of a rural WDN. We conclude the paper with a few remarks and an outline for the future work in Sec.~\ref{sec_conclusions}.

\section{Preliminaries} \label{sec_Prelim}

\subsection{Notations}
\begin{tabular}{cl}
$G$ & Graph \\
$E_G$ & Edge set of the graph $G$\\
$V_G$ & Vertex or node set of the graph $G$\\
$|X|$ & Cardinality of the set $X$ \\
$\mathbf{I}_G$ & Incidence matrix of graph $G$
\end{tabular}

\subsection{Graph preliminaries} \label{sec_graph_prelims}
For the following definitions, consider a graph, $G(E_G, V_G)$.


\subsubsection*{Path and directed path} A path in a graph is a sequence of edges which joins a sequence of distinct vertices. A directed path in a directed graph is a sequence of edges which joins a sequence of distinct vertices, with all the edges directed in the same direction.

\subsubsection*{Cycles and directed cycles} A cycle in a graph is a non-empty path in which the only repeated vertices are the first and last vertices. A directed cycle in a directed graph is a non-empty directed path in which the only repeated vertices are the first and last vertices. 

\subsubsection*{Tree} A tree is an undirected graph which has the following equivalent properties:
\begin{itemize}
\item $|E_G| = |V_G|-1$, that is the number of edges in the graph is one less than the number of vertices
\item There exists a unique path between a vertex and any other vertex
\item There are no cycles
\end{itemize}

\subsubsection*{Directed acyclic graph (DAG)} A finite directed graph with no directed cycles.

\subsubsection*{Polytree or a directed tree} A directed acyclic graph whose underlying undirected graph is a tree.

\subsubsection*{Connectivity or Connectedness} Two vertices $u$ and $v$ are called connected if the graph $G$ contains a path from $u$ to $v$. Otherwise, they are called disconnected.

\subsubsection*{Connected graph} An undirected graph is connected if it has at least one vertex and there is a path between every pair of vertices.

\subsubsection*{Connected Components} A connected component of an undirected graph  is a subgraph where there is a path from every node to every other node within the component.

\subsubsection*{Weakly Connected graph} A directed graph is called \emph{weakly connected} if replacing all of its directed edges with undirected edges produces a connected (undirected) graph.

\subsubsection*{Incidence matrix} The incidence matrix $\mathbf{I}_G$ of the directed graph $G$ with $|V|$ vertices and $|E|$ edges is defined as : \vspace{3pt}\\ 
\begin{tabular}{lcl}
$\mathbf{I}_G(i,j)$ & = & $+1$ if edge $j$ has its arrow leaving node $i$ \\
& & $-1$ if edge $j$ has its arrow entering node $i$ \\
& & $0$ if edge $j$ is not incident on node $i$
\end{tabular}
The incidence matrix is of size $|V| \times |E|$

\subsection{Assumptions}

\begin{assumption} \label{ass_wdn_dag}
	The WDN under study has a hierarchical structure of water flowing from source to multiple consumption zones. There are no loops that encompass multiple flow sensors meaning that a given flow sensor cannot be upstream of itself. This assumption is reasonable mostly in rural networks.  
\end{assumption}

\begin{assumption} \label{ass_static}
Throughout the paper, it is considered that the system under study is static. This is reasonable, since in a WDN, the sampling frequency is typically much lower than the actual dynamics of the system. Nevertheless, the same analysis could be extended to dynamical cases.
\end{assumption}

\section{Graph modelling} \label{sec_graph_modeling}



In this section, we illustrate the generic modelling of a WDN into a graph, accounting for sensor faults. For the sake of readability, this generic procedure is illustrated through the example of a WDN portion represented in Fig.~\ref{fig_tree3_schem}. The modelling process consists in three steps:
\begin{itemize} \itemsep -1pt
	\item[a)] Transform the sensor network into an equivalent electrical circuit (to provide an intuitive understanding of this procedure as well)
	\item[b)] Transform the equivalent circuit into a graph 
	\item[c)] Transform the measured information into residuals
\end{itemize}
which are detailed in the next section.

\subsection{Electrical Equivalent circuit}

\begin{figure}
\centering
\begin{circuitikz}[american,scale=0.45, every node/.style={scale=0.75}]
\draw (0,0) to [ammeter, *-*,  i=$\text{y}_1$, blue] ++ (4,0) -- (4,0) to [short,*-*] ++(2,0) node[label={above:$1$}]{};
\draw (6,0) to [short,*-,i=$ $] ++ (-3,-1.5) to [short,-] ++ (0,-1) to [vR=$R_{\mathcal{C}_1}$] ++(0,-2) -- ++ (0,-0.5) node[ground]{};
\draw (6,0) to [ammeter, *-*,  i=$\text{y}_2$, blue] ++ (4,0) node[label={above:$2$}]{}; 
\draw (10,0) to [short,*-,i=$ $] ++ (0,-0.5) to [vR=$R_{\mathcal{C}_2}$] ++(0,-2) -- ++ (0,-0.5) node[ground]{};
\draw (10,0) to [ammeter, *-*,  i=$\text{y}_3$, blue] ++ (6,0)node[label={above:$3$}]{} ;
\draw (16,0) to [short,*-,i=$ $] ++ (0,-0.5) to [vR=$R_{\mathcal{C}_3}$] ++(0,-2) -- ++ (0,-0.5) node[ground]{};
\draw (6,0) to [short, i=$ $] ++ (0,-5);
\draw (6,-5) to [ammeter, -*, i=$\text{y}_4$, blue] ++ (7,0) node[label={above:$4$}]{}
   to [short,*-,i=$ $] ++ (0,-0.5) to [vR=$R_{\mathcal{C}_4}$] ++ (0,-2) -- ++ (0,-0.5) node[ground]{};
\end{circuitikz}
\caption{Electrical equivalent circuit model of WDN}
\label{fig_elec_equiv_leaks_faults}
\end{figure}
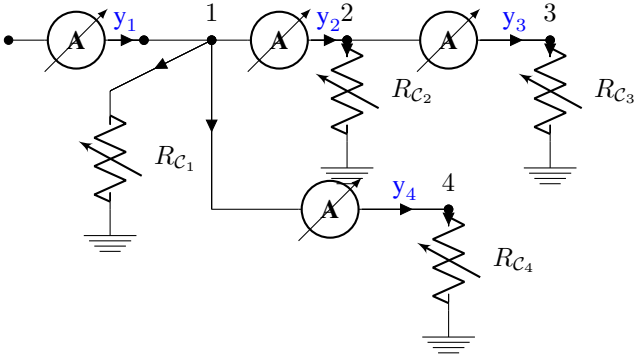

The approach takes a cue from the topological methods prominent in the electrical circuits (see for instance, \cite{narayanan1997submodular}). It is comparable to the classical work in \cite{carpentier_state_1991} with the addition of sensor fault in the mix.
Note that measuring flow in a water network is equivalent to measure the current in an electrical circuit. In this sense, any water consumption can be viewed as adding a current consuming resistance which might be known or in the case of leaks, unknown. 
Consequently, the WDN in Fig. \ref{fig_tree3_schem} is equivalent to the electrical circuit Fig.~\ref{fig_elec_equiv_leaks_faults} where the water consumption is represented by a Resistance (thanks to Assumption \ref{ass_static}) : the current flowing through a resistance $R_{\mathcal{C}_i}$ represents the entire consumption (regular consumption ${\mathcal{C}_i}$ and process fault like leaks ${\mathcal{L}_i}$) at a given point. Please note that dynamical study could be driven in the same way by adding other components like inductances or capacitors.

\begin{figure}[h]
\centering
\begin{minipage}{0.45\linewidth}
\begin{circuitikz}[american,scale=0.55, every node/.style={scale=0.65}]
\draw (0,0) to [ammeter, *-*, i=$y_i$, blue] ++ (4,0) -- (4,0) to [short,*-*] ++(2,0);
\draw (0.5,-3) to [american controlled current source, -o, i=$\mathcal{D}_i$, red] ++ (0,3);
\node at (0.5,-3) [ground]{};
\draw (4,3) to [american controlled current source,  i=$-\mathcal{D}_i$, red] ++ (0,-3);
\draw (4,3) to [short] ++ (0.5,0) -- ++ (0,0) node [ground]{};
\end{circuitikz}
\end{minipage}
\begin{minipage}{0.45\linewidth}
\begin{circuitikz}[american,scale=0.55, every node/.style={scale=0.65}]
\draw (0,0) to [ammeter, *-*, i=$y_i$, blue] ++ (6,0) -- (5.5,0) to [short,*-*] ++(0,-2);
\draw (5.5,-2) to [american controlled current source, -*, i=$\mathcal{D}_i$, red] ++ (-5,0) -- (0.5,-2) to [short,*-*]++(0,2);
\end{circuitikz}
\end{minipage}
\caption{Equivalent circuit for sensor faults}
\label{fig_eq_ckt_sensor_faults}
\end{figure}
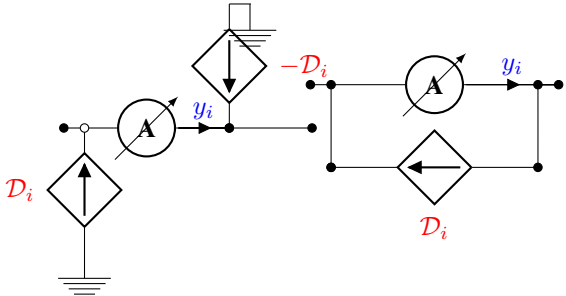

Process faults (leaks) display the same behaviour as regular consumption, but are anomalous in nature. Meanwhile, sensor faults (in current sensors) can be visualized as injection of current into and out of the paths on either side of the ammeter. This in turn can be viewed simply as a controlled current source in opposition to the ammeter (or the flow sensor), with the magnitude equal to that of the fault as shown in Fig.~\ref{fig_eq_ckt_sensor_faults}.


\subsection{Equivalent graph}
\begin{figure}
\centering
 \begin{tikzpicture}[scale=0.6, every node/.style={scale=0.75}]
    \node at (-4,0)[inner sep=3pt, draw, shape=circle] (nx){$X$};
    \node at (0,0)[inner sep=3pt, draw, shape=circle, fill=gray!30] (n1) {$1$};
    \node at (4,0)[inner sep=3pt, draw, shape=circle, fill=gray!30] (n2) {$2$};
    \node at (8,0)[inner sep=3pt, draw, shape=circle, fill=gray!30] (n3) {$3$};
    \node at (0,-3)[inner sep=3pt, draw, shape=circle, fill=gray!30] (n4) {$4$};     
    \node at (4,-3)[inner sep=3pt, draw, shape=circle] (n0){$0$};    
                
    \draw [->,blue] (nx) -- (n1) node[above, midway]{$\text{y}_1$};
    \draw [->,blue] (n1) -- (n2) node[above, midway]{$\text{y}_2$}; 
    \draw [->,blue] (n2) -- (n3) node[above, midway]{$\text{y}_3$}; 
    \draw [->,blue] (n1) -- (n4) node[left, midway]{$\text{y}_4$};
     

	\path (n1) edge[->, green!70!red] node [right,midway] {$\mathcal{L}_1 + \mathcal{C}_1$} (n0);
    \path (n2) edge[->, green!70!red] node [right,midway] {$\mathcal{L}_2+\mathcal{C}_2$} (n0);
    \path (n3) edge[->, green!70!red] node [right,midway] {$\mathcal{L}_3+\mathcal{C}_3$} (n0);
    \path (n4) edge[->, green!70!red] node [below,midway] {$\mathcal{L}_4+\mathcal{C}_4$} (n0);    
   
	\path (n1) edge[->,red, bend right=30] node [above] {$\mathcal{D}_{1}$} (nx);
	\path (n2) edge[->,red, bend right=30] node [above] {$\mathcal{D}_{2}$} (n1);
	\path (n3) edge[->,red, bend right=30] node [above] {$\mathcal{D}_{3}$} (n2);
	\path (n4) edge[->,red, bend left=40] node [left] {$\mathcal{D}_{4}$} (n1);
  \end{tikzpicture}
\caption{An equivalent graph model with leaks and faults added (The edge colouring is for the sake of readability)}
\label{fig_eg_graph_with_L_D}
\end{figure}
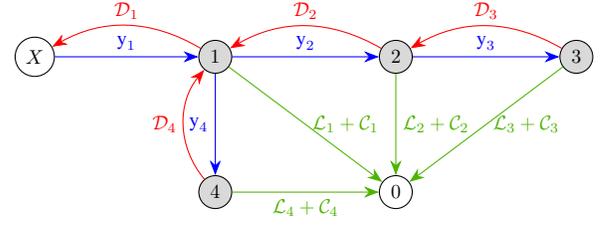

In order to make use of graph theory to obtain fault detectability conditions, the electrical circuit is turned into a graph. To ensure that the pipe junctions (such as those between $y_1$,$y_2$ and $y_4$ in this example) do not interfere in the graph representation, the sensors are represented by edges while the vertices represent physical consumptions points (DMAs). This representation is original in its form as typical graph representations would consider sensor measurements at the nodes and the system dynamics on the edges. The extra nodes of $X$ and $0$ refer to the source and sink nodes for the network respectively. This leads to the fact that water consumption at a particular DMA is an edge linking the respecting node and the sink node. Finally the sensor faults are represented by edges oriented from a sensor downstream node towards the sensor upstream node. This results, in our example, the graph represented in Fig. \ref{fig_eg_graph_with_L_D}.

\subsection{Turning the measurements into residuals}

The ultimate aim of the graph formulation remains to estimate leaks and faults. One last remaining problem in the obtained graph is the inability to separate the regular consumptions $\mathcal{C}_{i}$ from the system faults $\mathcal{L}_{i}$. This can be done by simply subtracting any model-predicted estimate of the consumption  $\mathcal{M}_i$  from the measured information  $\text{y}_i$ leading to residuals $\mathcal{E}_i = \text{y}_i - \mathcal{M}_i$. $\mathcal{M}_i$ can be issued from any kind of model (data-based models, system physical models, \emph{a priori} information, etc.). Should the $\mathcal{M}_i$ be precise enough, subtracting the prediction from the measurement is equivalent to removing all regular consumptions terms from the graph.

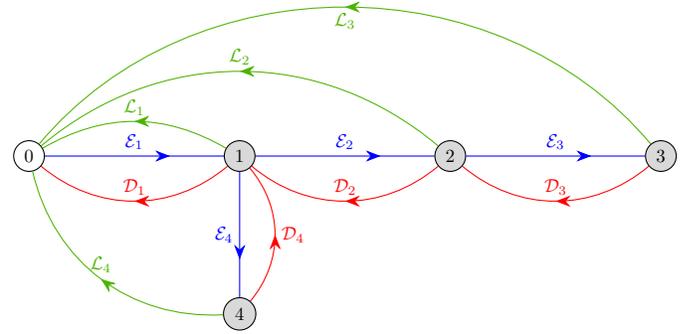
\begin{figure}
\centering
 \begin{tikzpicture}[scale=0.7, every node/.style={scale=0.7}]
    \node at (-4,0)[inner sep=3pt, draw, shape=circle] (nx){$0$};
    \node at (0,0)[inner sep=3pt, draw, shape=circle, fill=gray!30] (n1) {$1$};
    \node at (4,0)[inner sep=3pt, draw, shape=circle, fill=gray!30] (n2) {$2$};
    \node at (8,0)[inner sep=3pt, draw, shape=circle, fill=gray!30] (n3) {$3$};
    \node at (0,-3)[draw, shape=circle, fill=gray!30] (n4) {$4$};     
    \draw [->-=0.7,blue] (nx) -- (n1) node[above, midway]{$\mathcal{E}_1$};
    \draw [->-=0.7,blue] (n1) -- (n2) node[above, midway]{$\mathcal{E}_2$}; 
    \draw [->-=0.7,blue] (n2) -- (n3) node[above, midway]{$\mathcal{E}_3$}; 
    \draw [->-=0.7,blue] (n1) -- (n4) node[left, midway]{$\mathcal{E}_4$};
    \path (n1) edge[->-=0.5,green!70!red, bend right=30] node [above,midway] {$\mathcal{L}_{1}$} (nx);
    \path (n2) edge[->-=0.5,green!70!red, bend right=40] node [above,midway] {$\mathcal{L}_{2}$} (nx);
    \path (n3) edge[->-=0.5,green!70!red, bend right=50] node [below,midway] {$\mathcal{L}_{3}$} (nx);
    \path (n4) edge[->-=0.5,green!70!red, bend left=40] node [above,midway] {$\mathcal{L}_{4}$} (nx);
	\path (n1) edge[->-=0.5,red, bend left=40] node [above] {$\mathcal{D}_{1}$} (nx);
	\path (n2) edge[->-=0.5,red, bend left=40] node [above] {$\mathcal{D}_{2}$} (n1);
	\path (n3) edge[->-=0.5,red, bend left=40] node [above] {$\mathcal{D}_{3}$} (n2);
	\path (n4) edge[->-=0.5,red, bend right=40] node [right] {$\mathcal{D}_{4}$} (n1);
  \end{tikzpicture}
\caption{Graph model ($G_{\mathcal{FR}}$) for analysis}
\label{fig_topological_graph_residues_leaks_faults}
\end{figure}

Finally, it is intuitive to notice that the source node $X$ provides exactly the sum of consumptions arriving at the sink node $0$. therefore these nodes are redundant and can be combined without affecting the analysis on the other nodes.  We refer to the $0$ node as the reference node/vertex in this paper. In the electrical circuit paradigm, this operation is equivalent to short-circuiting a voltage source. A  more formal analysis of this merging for topological analysis can be referred from Theorem 6.3.1 in \cite{narayanan1997submodular}. The resulting graph corresponding to that of our example is presented in Fig.~\ref{fig_topological_graph_residues_leaks_faults}.

In the rest of the paper, the resulting graph (such as represented in Fig.~\ref{fig_topological_graph_residues_leaks_faults} in our example), containing both the residual and fault edges, is denoted as $G_{\mathcal{FR}}$. It is important to note that the residual edges (in blue) represent measured information and a known quantity, while the fault edges (in red and green) represent unknown quantities. \\

\noindent Furthermore the following notations will also be used:
\begin{itemize} \itemsep -1pt
	\item the residual graph obtained from  $G_{\mathcal{FR}}$  by removing the fault edges is denoted $G_{\mathcal{R}}$ 
	\item the fault graph from  $G_{\mathcal{FR}}$  by removing the residual edges is denoted $G_{\mathcal{F}}$ 
\end{itemize}

\subsection{Properties of the Obtained Graphs}

Firstly, Assumption \ref{ass_wdn_dag} can be reformulated as the following assumption
\subsubsection*{Assumption \ref{ass_wdn_dag}b}
The equivalent graph of the underlying water distribution network  $G_{\mathcal{R}}$ is a directed acyclic graph (DAG).

Moreover, We note the following properties of these graphs:
\begin{itemize} \itemsep -1pt
\item $G_{\mathcal{R}}$ is a directed tree.
\item All the process fault (leak) edges are incident on the reference node $0$.
\item The sensor fault edges are in opposition to the corresponding residue edges. 
\item All the three graphs have the same set of nodes. We would refer to their vertex set as $V_{G_\mathcal{FR}}$.
\item There is at most one outgoing sensor fault edge per node. 
\item There is at most one outgoing process fault edge per node.
\end{itemize}

\begin{lemma} \label{lemma_types_of_CC_after_residue_removal}
The following are some properties of $G_\mathcal{F}$
\begin{enumerate}
\item $G_\mathcal{F}$ contains connected components that involve the nodes of $G_{\mathcal{FR}}$. Each node in $G_{\mathcal{FR}}$ belongs to one and only one of the below components.
\item The graph $G_\mathcal{F}$ may consist of the following types of components:
	\begin{enumerate}
	\item $UC_i$: $n_{UC}$ unconnected nodes \label{type1_cc}
	\item $G_{\mathcal{F}_{S_i}}$: $n_S$ weakly connected components with only sensor fault edges \label{type2_cc}
	\item $G_{\mathcal{F}_{PS}}$: $1$ weakly connected component which has the reference node and possibly the process faults and sensor faults edges	\label{type3_cc}
	\end{enumerate}
\end{enumerate}
\end{lemma}
\paragraph*{Proof} 
\begin{enumerate}
\item $G_\mathcal{F}$ is obtained by removing the residue edges on $G_{\mathcal{FR}}$. The connected components are formed based on the fault edges left out. Hence a node can be part of only one of the connected components in $G_\mathcal{F}$.
\item Since $G_\mathcal{F}$ contains two types of edges, the connected components of types \ref{type1_cc} and \ref{type2_cc} are a consequence of a simple combination. For the case of the component type \ref{type3_cc}, note the following:
	\begin{itemize}
	\item By construction, all the process fault edges are incident on the reference node  (if present). Hence there can be only one component that contains all process fault edges.
	\item If a sensor fault edge is incident on the reference node and there are process fault edges in $G_\mathcal{F}$, then this sensor fault edge would be part of this component.
	\item Let's denote the set of nodes with an outgoing process fault edge as $V_P$. These nodes are present in the component type \ref{type3_cc} as pointed out earlier. Further, it is possible that a node (or several nodes) is connected to a node in $V_P$ through sensor fault edges. Hence these nodes and the sensor fault edges would also belong to the connected component of type \ref{type3_cc}.
	\end{itemize}
\end{enumerate}
\subsection{Example}

\begin{figure}
	\centering
	\begin{subfigure}[$UC_1$]
	\centering
	  \begin{tikzpicture}[scale=0.65, every node/.style={scale=0.75}]
	    \node at (8,-3)[inner sep=3pt, draw, shape=circle, fill=gray!30] (n6) {$6$};
	 \end{tikzpicture}
	\label{a}
	\end{subfigure}
	$\qquad$
	\begin{subfigure}[$G_{\mathcal{F}_{S_1}}$]
		\centering
		  \begin{tikzpicture}[scale=0.65, every node/.style={scale=0.75}]
		    \node at (0,0)[inner sep=3pt, draw, shape=circle, fill=gray!30] (n1) {$1$};
		    \node at (0,-3)[inner sep=3pt, draw, shape=circle, fill=gray!30] (n4) {$4$};     
			\path (n4) edge[->-=0.5,red, bend right=40] node [right] {$\mathcal{D}_{4}$} (n1);	
		 \end{tikzpicture}
		 		 \label{b}
	\end{subfigure}
	\begin{subfigure}[$G_{\mathcal{F}_{PS}}$]
		\centering
		  \begin{tikzpicture}[scale=0.65, every node/.style={scale=0.75}]
		     \node at (2,1.5)[inner sep=3pt, draw, shape=circle] (nx){$0$};
		     \node at (4,0)[inner sep=3pt, draw, shape=circle, fill=gray!30] (n2) {$2$};
		     \node at (8,0)[inner sep=3pt, draw, shape=circle, fill=gray!30] (n3) {$3$};    
		     \node at (4,-3)[inner sep=3pt, draw, shape=circle, fill=gray!30] (n5) {$5$};
		     \path (n3) edge[->-=0.5,green!70!red, bend right=25] node [below,midway] {$\mathcal{L}_{3}$} (nx);
		     \path (n5) edge[->-=0.5,green!70!red, bend left=20] node [left,midway] {$\mathcal{L}_{5}$} (nx);
		 	\path (n3) edge[->-=0.5,red, bend left=30] node [above] {$\mathcal{D}_{3}$} (n2);
		 	\path (n5) edge[->-=0.5,red, bend right=40] node [right] {$\mathcal{D}_{5}$} (n2);
		 \end{tikzpicture}
		 \label{c}
	\end{subfigure}	
	\caption{Connected Components in $G_\mathcal{F}$ }
	\label{fig_eg1_cc}
	\end{figure}
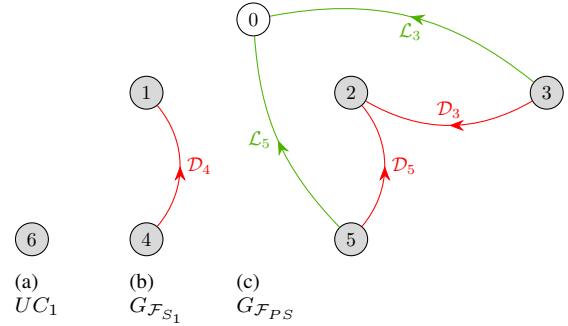

As an example, consider the equivalent graph of a WDN with leaks and sensor faults as given in Fig.~\ref{fig_eg1_gfr}.  The equivalent $G_\mathcal{F}$ graph after removing the residue edges, composed of $\mathcal{F}_{S_1}$, $UC_1$ and $G_{\mathcal{F}_{PS}}$ is given in Fig.~\ref{fig_eg1_cc}. 
\begin{figure}
\centering
 \begin{tikzpicture}[scale=0.65, every node/.style={scale=0.75}]
    \node at (-4,0)[inner sep=3pt, draw, shape=circle] (nx){$0$};
    \node at (0,0)[inner sep=3pt, draw, shape=circle, fill=gray!30] (n1) {$1$};
    \node at (4,0)[inner sep=3pt, draw, shape=circle, fill=gray!30] (n2) {$2$};
    \node at (8,0)[inner sep=3pt, draw, shape=circle, fill=gray!30] (n3) {$3$};
    \node at (0,-3)[inner sep=3pt, draw, shape=circle, fill=gray!30] (n4) {$4$};     
    \node at (4,-3)[inner sep=3pt, draw, shape=circle, fill=gray!30] (n5) {$5$};
    \node at (8,-3)[inner sep=3pt, draw, shape=circle, fill=gray!30] (n6) {$6$};
    \draw [->-=0.7,blue] (nx) -- (n1) node[above, midway]{$\mathcal{E}_1$};
    \draw [->-=0.7,blue] (n1) -- (n2) node[above, midway]{$\mathcal{E}_2$}; 
    \draw [->-=0.7,blue] (n2) -- (n3) node[above, midway]{$\mathcal{E}_3$}; 
    \draw [->-=0.7,blue] (n1) -- (n4) node[left, midway]{$\mathcal{E}_4$};
    \draw [->-=0.7,blue] (n2) -- (n5) node[left, midway]{$\mathcal{E}_5$}; 
    \draw [->-=0.7,blue] (n3) -- (n6) node[left, midway]{$\mathcal{E}_6$};    
    \path (n3) edge[->-=0.5,green!70!red, bend right=25] node [below,midway] {$\mathcal{L}_{3}$} (nx);
    \path (n5) edge[->-=0.5,green!70!red, bend left=45] node [below,midway] {$\mathcal{L}_{5}$} (nx);
	\path (n3) edge[->-=0.5,red, bend left=30] node [above] {$\mathcal{D}_{3}$} (n2);
	\path (n4) edge[->-=0.5,red, bend right=40] node [right] {$\mathcal{D}_{4}$} (n1);
	\path (n5) edge[->-=0.5,red, bend right=40] node [right] {$\mathcal{D}_{5}$} (n2);
 \end{tikzpicture}
 \caption{Example graph $G_\mathcal{FR}$}
\label{fig_eg1_gfr}
\end{figure}
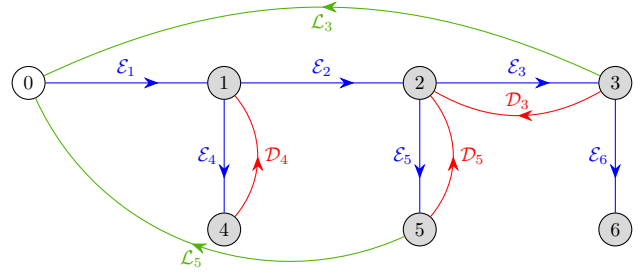

The next section defines the problem of detectability for a given graph.

\section{Fault detectability} \label{sec_problem_formulation}
\subsection{Problem formulation}
In this section, we formulate the simultaneous detectability problem, which implicitly tackles the distinguishability of two types of faults, the process faults and the sensor faults. If we only consider process faults, the detectability property is equivalent to the topological observability problem treated in the literature (\hspace{1sp}\cite{carpentier_state_1991}, \cite{diaz2017topological}). However, the presence of sensor faults necessitates a reformulation of the topological observability problem. Note that, similar to the topological observability, simultaneous detectability refers to a generic property. Hence practical detectability can breakdown for particular fault values.

In practice, fault detection is a two-step operation. One estimates the variables which are monitored in the first step. In the second step, this estimation is compared with a threshold, and if it exceeds the threshold, a fault is \textit{detected}. The threshold option, however, is unique to each problem, and can not be generalized. Hence, we concentrate on the first step, the ability to estimate the faults, and use it to validate the underlying system's detectability property. This transforms the problem of detectability into the ability to estimate the unknown variables (the process and sensor faults in our case).

This leads to the following definition for the graph $G_{\mathcal{FR}}$ that contains the known residues and the unknown faults.

\begin{definition}
A graph is termed \emph{detectable} if all the unknown faults in it are solvable.
\end{definition}

Topological methods concern performing nodal analysis on a graph's nodes \cite{narayanan1997submodular} and leads to a set of linear equations. Hence, the detectability of the unknowns in a graph can be captured through the solvability of the unknowns in these linear equations. One computationally convenient way to capture these linear equations is using the graph's incidence matrix \cite{harary1967graph}. If we refer to $X$ as the set of all variables of interest in a graph ($\mathcal{E}_i$, $\mathcal{L}_i$, $\mathcal{D}_i$ in case of the graph ${G_{\mathcal{FR}}}$), then the set of nodal equations (for a given configuration of process and sensor faults) is given by:
\begin{align}\label{eq_inc_matrix_eqn_1}
\mathbf{I}_{G_\mathcal{FR}} X = \mathbf{0}
\end{align}
It is evident that the solvability (and hence the detectability) of the faults $\mathcal{L}_i$, $\mathcal{D}_i$ can be inferred from \eqref{eq_inc_matrix_eqn_1}. However, to glean out the solvability from this equation, we need to focus on a part of the matrix $\mathbf{I}_{G_\mathcal{FR}}$. We do this by grouping the known and unknown variables as,
\begin{align} \label{eq_inc_matrix_eqn_2}
\mathbf{A}_{G_\mathcal{F}} X_\mathcal{F} = \mathbf{B}_{G_\mathcal{R}} X_\mathcal{R}
\end{align}
where $X_\mathcal{F}$ is a vector that contains all the faults and is of length $|E_{G_\mathcal{F}}|$ and $X_R$ contains the residues and of length $|E_{G_\mathcal{R}}|$. Here $E_{G_\mathcal{F}}$ and $E_{G_\mathcal{R}}$ refer to the edge set of the graphs $G_\mathcal{F}$ and $G_\mathcal{R}$ respectively. The matrix $\mathbf{A}_{G_\mathcal{F}}$ is of dimension $|V_{G_\mathcal{FR}}| \times |E_{G_\mathcal{F}}|$ and $\mathbf{B}_{G_\mathcal{R}}$ of dimension $|V_{G_\mathcal{FR}}| \times |E_{G_\mathcal{R}}|$. One can note that
$\mathbf{A}_{G_\mathcal{F}} $ and $\mathbf{B}_{G_\mathcal{R}} $ are the incidence matrices of the graphs $G_\mathcal{F}$ and $G_\mathcal{R}$ respectively (with appropriately modified sign conventions). It is hence possible to characterize the solvability and hence the detectability through the analysis of the matrix ranks in \eqref{eq_inc_matrix_eqn_2}. 

Our interest is to understand if this can be achieved by directly analyzing the underlying graphs, in particular, $G_\mathcal{F}$.

\subsection*{Problem statement and solution approach}
At time $t$, given a static sensor network, its associated residuals $\mathcal{E}_i$ represented by its graph $G_{\mathcal{R}}$ that satisfies the Assumption \ref{ass_wdn_dag}b, a set of possible faults (process or sensor) represented by a graph $G_{\mathcal{F}}$, what are the conditions under which $G_{\mathcal{F}}$ is detectable?

Since ${G_\mathcal{F}}$ represents $N=n_{UC} + n_S +1$ subgraphs (\textit{Lemma \ref{lemma_types_of_CC_after_residue_removal}}). The global detectability problem of ${G_\mathcal{F}}$ in \eqref{eq_inc_matrix_eqn_2} is equivalent to the solvability of each component of ${G_\mathcal{F}}$ and therefore, can be split in $n_S +1$ subproblems (Unconnected components do not contain any unknowns):

\begin{equation}
	\begin{cases} \label{eq_inc_matrix_eqn_4}
	\mathbf{A}_{G_{\mathcal{F}_{PS}}} X_{\mathcal{F}_{PS}} = \mathbf{B}_{G_{\mathcal{F}_{PS}}} X_{\mathcal{R}_{PS}}\\
	\mathbf{A}_{G_{\mathcal{F}_{S_i}}} X_{\mathcal{F}_{S_i}} = \mathbf{B}_{G_{\mathcal{F}_{S_i}}} X_{\mathcal{R}_{S_i}}  \quad \forall i \in [1,\ldots,n_S]
	\end{cases}
\end{equation}

 Where 
 \begin{align}
 	&\mathbf{A}_{G_{\mathcal{F}_{PS}}} \in \mathbb{R}^{|V_{G_{\mathcal{F}_{PS}}}| \times |E_{G_{\mathcal{F}_{PS}}}|} \notag\\
	&X_{G_{\mathcal{F}_{PS}}} \in \mathbb{R}^{|E_{G_{\mathcal{F}_{PS}}}|\times 1} \notag\\
	&\mathbf{B}_{\mathcal{F}_{PS}} \in \mathbb{R}^{|V_{G_{\mathcal{F}_{PS}}}| \times (|V_{G_{\mathcal{F}_{PS}}}|-1)}\notag\\
	&X_{\mathcal{R}_{PS}} \in \mathbb{R}^{(|V_{G_{\mathcal{F}_{PS}}}|-1) \times 1} \notag	
 \end{align}
 and
 \begin{align}\notag
 	&\mathbf{A}_{G_{\mathcal{F}_{S_i}}} \in \mathbb{R}^{|V_{G_{\mathcal{F}_{S_i}}}| \times |E_{G_{\mathcal{F}_{S_i}}}|}, 
	\quad X_{\mathcal{F}_{PS}} \in \mathbb{R}^{|E_{G_{\mathcal{F}_{S_i}}}|\times 1} \notag \\
	&\mathbf{B}_{G_{\mathcal{F}_{S_i}}} \in \mathbb{R}^{|V_{G_{\mathcal{F}_{S_i}}}| \times (|V_{G_{\mathcal{F}_{S_i}}}|-1)} \notag\\
	&X_{\mathcal{F}_{S_i}} \in \mathbb{R}^{(|V_{G_{\mathcal{F}_{S_i}}}|-1) \times 1} \notag\\
	&\forall i \in [1,\ldots,n_S] \notag
 \end{align}
 
 For readability's sake each subproblem is denoted in the rest of the paper as 
 \begin{equation} \label{eq_inc_matrix_eqn_5}
 	\mathbf{A}_{G_{\mathcal{F}_{j}}} X_{\mathcal{F}_{j}} = \mathbf{B}_{G_{\mathcal{F}_{j}}} X_{\mathcal{R}_{j}} \forall i \in [1,\ldots,n_S+1]
 \end{equation}
 
In order to present the detectability results, some preliminaries are required.

\subsection{Preliminaries}

\begin{lemma}[\hspace{1sp}\cite{harary1967graph}, \cite{deo2017graph}] \label{lemma_incidence_matrix_rank}
If $I_G$ is an incidence matrix of a connected graph $G$  with $n$ vertices, the rank of $I_G$ is $n-1$. If there are $m$ connected components, then the rank is $n-m$.
\end{lemma}

\begin{definition}[Reduced incidence matrix] The reduced incidence matrix of a connected graph with $|V|$ nodes and $|E|$ edges is any $(|V|-1) \times |E|$ block of its incidence matrix.
\end{definition}
\begin{lemma}[\hspace{1sp}\cite{deo2017graph}] \label{lemma_reduced_incidence_matrix}
The reduced incidence matrix of a directed tree is nonsingular.
\end{lemma}

\begin{remark}
The above properties of the incidence matrix informs us that the set of linear equations represented by \eqref{eq_inc_matrix_eqn_1} is always consistent. That is, the equations lead to either a unique or infinitely many solutions.
\end{remark}

\subsection{Main result}

\begin{theorem} \label{thm_detectability}
$G_\mathcal{F}$ is detectable if and only if every connected component of $G_\mathcal{F}$  is a directed tree.
\end{theorem}
\paragraph*{Proof} 
\paragraph*{Sufficiency}
If a given connected component is a directed tree, the following are the consequences:
\begin{itemize}
\item The number of unknowns (faults) $|X_{G_{\mathcal{F}_j}}|$ in each of the connected component $G_{\mathcal{F}_j}$, is equal to the number of edges $E_{G_{\mathcal{F}_j}} = |V_{G_{\mathcal{F}_j}}-1|$.
\item The reduced incidence matrices  of $G_{\mathcal{F}_j}$  has a rank $|V_{G_{\mathcal{F}_j}}-1|$ (as per \textit{Lemma \ref{lemma_reduced_incidence_matrix}}).
\end{itemize}
This means that, for each component of $G_{\mathcal{F}}$ there are equal number of independent rows (and hence nodal equations) as the number of faults. Hence the faults are solvable, every $G_{\mathcal{F}_j}$ is detectable and $G_{\mathcal{F}}$ is detectable. 

\paragraph*{Necessity}

By assuming that $G_{\mathcal{F}}$ and consequently all its components $G_{\mathcal{F}_j}$ are detectable, then we have :
\begin{align}\label{eq_theorem_necessity_1}
|E_{G_{\mathcal{F}_{j}}}| \leq |V_{G_{\mathcal{F}_{j}}}|-1
\end{align}
This is a consequence of \textit{Lemma \ref{lemma_incidence_matrix_rank}} and that all the matrices $\mathbf{A}_{G_{\mathcal{F}_{j}}}$ have a rank $|V_{G_{\mathcal{F}_{PS}}}|-1$.
The following are the known properties of the connected component $F_{PS}$,
\begin{itemize}
\item In every ${G_{\mathcal{F}_{j}}}$, there is one node that has no outgoing edges.
\item Every other node in ${G_{\mathcal{F}_{j}}}$ has at least one outgoing or incoming edge.
\end{itemize}
This means, the number of edges is at least one less than the number of nodes in ${G_{\mathcal{F}_{j}}}$. Hence,
\begin{align}\label{eq_theorem_necessity_2}
|E_{G_{\mathcal{F}_{j}}}| \geq |V_{G_{\mathcal{F}_{j}}}|-1
\end{align}
By putting together \eqref{eq_theorem_necessity_1} and \eqref{eq_theorem_necessity_2}, we get
\begin{align}\label{eq_theorem_necessity_tree}
|E_{G_{\mathcal{F}_{j}}}| = &|V_{G_{\mathcal{F}_{j}}}|-1\  \Rightarrow \  {G_{\mathcal{F}_{j}}} \text{are Directed Trees} \notag \\ &\Rightarrow \  {G_{\mathcal{F}}} \text{ contains only Directed Trees}
\end{align}%
\noindent This completes the proof. $\blacksquare$

\subsection{Discussion}

\begin{remark} [Computational complexity] 
Theorem~\ref{thm_detectability}  states that checking if ${G_{\mathcal{F}}}$ is a directed tree is sufficient to assess the detectability of ${G_{\mathcal{F}}}$.  The complexity of such an algorithm is $\mathcal{O}(n)$. In case of a depth-first search approach, the worst-case scenario is with $n=|V_{G_{\mathcal{F}}}|+|E_{G_{\mathcal{F}}}|$. On the other hand, to verify detectability using the matrix rank of A would be of the order of $\mathcal{O}(n^3)$ when using Gaussian elimination \cite{watkins2004fundamentals}. Some approaches offer better complexity, but all of them are polynomial-time algorithms.

	Hence the graphical approach offers a faster way to verify detectability for large scale systems or if one has to perform this operation several times.
\end{remark}

\begin{remark}
	Since $G_{{\mathcal{F}_{S_i}}}\ (\forall i)$ are directed trees by construction, Theorem~\ref{thm_detectability} illustrates that $G_\mathcal{F}$ can be undetectable if and only if $G_\mathcal{F_{PS}}$ is undetectable. Hence it is sufficient to verify only the detectability of this connected component. More generally, one tests whether the connected component containing the reference node is a directed tree or not. This is summarized in the Algo.~\ref{alg_detectability_global}.
\end{remark}

\begin{algorithm}[ht]
\begin{algorithmic}[1]
\STATE Given: $G_{\mathcal{F}}$
\STATE Obtain the set of connected   components $G_{\mathcal{F}_{PS}}$  and $G_{\mathcal{F}_{S_i}}\ (i=1,..n_S)$  in $G_{\mathcal{F}}$
\IF{$G_{\mathcal{F}_{PS}}$ is a directed tree}
	\STATE Detectable
\ELSE
	\STATE Not detectable
\ENDIF
\end{algorithmic}
\caption{An algorithm for the detectability of faults}
\label{alg_detectability_global}
\end{algorithm}

\begin{remark}
If a given $G_\mathcal{F}$ is declared undetectable, one can still extract a part of the detectable faults. For instance, all connected components of type $G_{\mathcal{F}_{S_i}}$ are detectable. Further, one can remove vertices in  $G_{\mathcal{F}_{PS}}$ to determine which sensor causes undetectability.
\end{remark}

\begin{corollary} \label{coll_theorem_1} If all vertices in the graph $G_\mathcal{FR}$ contain at most one sensor fault or one process fault simultaneously, then $G_\mathcal{F}$ is detectable.
\paragraph*{Proof}
 If all vertices in the graph $G_\mathcal{FR}$ contain at most one sensor fault or one process fault, then $G_{\mathcal{F}_{PS}}$ is a directed tree by construction.
\end{corollary}

\begin{remark}
$G_\mathcal{F}$ presented in Fig. \ref{fig_eg1_cc} is not detectable since $G_{\mathcal{F}_{PS}}$  is not a directed tree.
\end{remark}

\section{Application: Leak and Sensor fault estimation in a WDN} \label{sec_leak_detection}

This section presents the application of the above results in the context of leak estimation in a WDN. This work derives motivation from the necessity of an efficient monitoring of a water network (to avoid leaks) through a more intuitive and explainable estimation. 

In this problem, the graph ${G_\mathcal{R}}$ is known (and hence the structure of the sensor network and the water flow directions are known). If the true structure  of ${G_\mathcal{F}}$ was \emph{a priori} known and detectable, then the faults $ X_\mathcal{F}$ of ${ G_\mathcal{F}}$ could be estimated using simple least squares on a residues data series  $X_\mathcal{R}$: 
\begin{equation}\label{est_prob_L2}
	{\hat X_\mathcal{F}}=\min_{X_\mathcal{F}} {\lVert \mathbf{A}_{G_\mathcal{F}} X_\mathcal{F} - \mathbf{B}_{G_\mathcal{R}} X_\mathcal{R} \rVert^2 }
\end{equation}

However, it is not known \emph{a priori} which faults are present on the network at any given time. Consequently it is also not known what is the true ${G_\mathcal{F}}$ and whether it is detectable or not. Hence, the estimation problem can be summarized as: Given ${G_\mathcal{R}}$ and the associated sensor data, is it possible to find ${\hat G_\mathcal{F}}$, an estimate of the true fault graph ${G_\mathcal{F}}$, along with the values of the associated faults $ X_\mathcal{F}$  contained in it? 

Since ${\hat G_\mathcal{F}}$ is unknown, one strategy would be to assume a general underlying fault structure ${{\mathcal{G}_\mathcal{F}}}$ where every node has a leak and every sensor is faulty. And all these leaks and faults are stacked in a vector ${\mathcal{X}_\mathcal{F}}$. For example, for the WDN as in Fig. \ref{fig_tree3_schem}, the set ${{\mathcal{G}_\mathcal{F}}}$ is shown in Fig. \ref{fig_topological_graph_residues_leaks_faults}. Naturally, this problem is over-parameterized since $|E_{G_\mathcal{F}|}=2\times (|V_{G_\mathcal{F}}|-1)$. This consequently needs some regularization. 

In order to regularize this problem, the following assumption is made: the most likely structure of ${\hat G_\mathcal{F}}$ is the one fitting the data while requiring the least number of variables. This is an $\ell_2-\ell_0$ problem which can be relaxed using the well-known $\ell_2-\ell_1$ LASSO regularization method. One can add temporal information, fault shapes, etc., to be included in the optimization problem. However, the aim of this section is not to formulate an optimization problem that can provide the best estimate, but to illustrate the use of the underlying graph structure in the estimation that can provide a more explainable estimation. Consequently, the proposed optimization scheme is not discussed in this paper. It can only be said that the assumption is safe in a WDN where leaks and sensor faults are repaired when detected.

Further, one can add physical constraints such as the positivity of leaks (water can only flow out of the network), leading to the final optimization criterion as:
\begin{align}\label{est_problem}
	{\hat X_\mathcal{F}}&=\min_{{\mathcal{X}_\mathcal{F}}} {\lVert \mathbf{A}_{{\mathcal{G}_\mathcal{F}}} \mathcal{X}_\mathcal{F}\!-\!\mathbf{B}_{G_\mathcal{R}} X_\mathcal{R} \rVert^2 }
	\!+\!\lambda  \Vert  \mathcal{X}_\mathcal{F} \Vert_1 \notag\\
		& s.t \quad \mathcal{L}_i>0 \quad \forall i \in [1,\ldots,|E_{G_\mathcal{R}}|-1] 
\end{align}

where $\lambda$ is the \emph{so-called} regularization hyper-parameter. This problem can be written as a quadratic minimization problem under linear matrix inequality constraints and using quadratic programming. 

\begin{remark}
A note on the optimization formulation. The detectability of ${{G}_\mathcal{F}}$ does not guarantee the convergence to the true solution (as the detectability property is bound to a given structure and not the solution).  Therefore, it is possible that this problem has multiple solutions where different structures minimize \eqref{est_problem}. In practice, there are several reasons possible: the true structure ${{G}_\mathcal{F}}$ is not a solution of \eqref{est_problem}, or  it does not fit the regularization assumption (minimizing the number of faults), or perhaps that it is not be detectable. Consequently it can only be claimed that ${\hat{G}_\mathcal{F}}$ is guaranteed to lie within the set of solutions of \eqref{est_problem} if and only if ${{G}_\mathcal{F}}$ is detectable \emph{and} its fault norm is minimal. 
\end{remark}

Even under these conditions, solving this problem using quadratic programming presents several drawbacks:
\begin{itemize}
\item If \eqref{est_problem} has more than one solutions, a gradient descent algorithm will stop at one of the minima and provide only one solution (even if the proposed ${\hat G_\mathcal{F}}$ is detectable). 
 \item Gradient descent algorithm for quadratic programming under linear constraints can be time consuming (considering one data point every 15 minutes, then estimating the leaks over a year of data corresponds to 35040 estimation problems in the form of \eqref{est_problem}).
 	\item The requirement for tuning the hyper-parameter  $\lambda$.
\end{itemize} 
In this paper, the quadratic programming implementation of the optimization problem in  \eqref{est_problem} is referred to as \textit{QP-Lasso}. The next section depicts how the optimization problem in \eqref{est_problem} can be constrained using the proposed Theorem \ref{thm_detectability}.

\subsection{Proposed algorithm}

Thanks to Theorem \ref{thm_detectability}, it is possible to \emph{a priori} determine a finite set ${\bar{\mathcal{G}}_\mathcal{F}}$ of all possible \emph{detectable} graphs ${G_\mathcal{F}}$. Doing so presents many advantages:
\begin{itemize}
	\item Since ${\hat G_\mathcal{F}}$ is searched within a finite set ${\bar{\mathcal{G}}_\mathcal{F}}$, it reduces the number of possible solutions in order to speed up the process. Furthermore, detectable graphs represent full rank equation systems, for which it is possible to use directly the optimization scheme \eqref{est_prob_L2}. Consequently, simple analytic least-squares method can be used in order to further computationally speed up the process.
	\item Since ${\bar{\mathcal{G}}_\mathcal{F}}$ is finite, it is possible to compute all the solutions of \eqref{est_problem}.  It is therefore possible to hand out a set of possible solutions (computed as the minimum-maximum range of the given faults among all detectable solutions found) in which the true underlying graph  ${G_\mathcal{F}}$ is guaranteed to lie, if it is detectable. 
\end{itemize}
Hence, the global optimization problem at hand in \eqref{est_problem} can be split into several steps presented in Algorithm~\ref{alg_constraint}. One could argue that the computation time gained from using multiple well-posed problems in PART 2, instead of a regularized global problem \eqref{est_problem} may be counteracted by the computation time of PART 1 (determining $\bar{\mathcal{G}}_\mathcal{F}$). However, while the leak estimation is an online problem solved on a fix sensor network, the possibly time consuming PART 1 will be computed only once. This online/offline splitting is not possible when using the {QP-Lasso} approach.

\begin{algorithm}
\begin{algorithmic}[1]
\STATE Input: a graph $G_\mathcal{R}$ and sensor data
\STATE PART 1: Determine the set $\bar{\mathcal{G}}_\mathcal{F}$ of  all detectable  $G_{F}$ containing $|V_{G_\mathcal{R}}|-1$ edges by repeated application of Algo.~\ref{alg_detectability_global}. \label{step_algo2_offline}
 \STATE PART 2: 
 \FOR{every ${G^j}_\mathcal{F} \in \bar{\mathcal{G}}_\mathcal{F} $ }
 	\STATE Solve $\mathbf{A}_{G^j_\mathcal{F}} X^j_\mathcal{F}=\mathbf{B}_{G^j_\mathcal{R}} X^j_\mathcal{R}$ (or  \eqref{est_prob_L2} if the leak are identified from several data points)
 	\STATE Test if all Leaks in $X^j_\mathcal{F}$ are positive. If so, consider the model \emph{valid}. If not, the identified model is declared \emph{invalid}
 	\STATE Compute the $\ell_1$ norm $|X^j_\mathcal{F}|$
 \ENDFOR
 \STATE Output: All \emph{valid} ${G^j}_\mathcal{F}$ presenting the smallest value for $|X^j_\mathcal{F}|$
\end{algorithmic}
\caption{An algorithm to identify leaks in a WDN}
\label{alg_constraint}
\end{algorithm}

In the next section, Algorithm~\ref{alg_constraint} is compared to QP-Lasso implementation in \eqref{est_problem} on a simulated dataset from  the network depicted in Fig. \ref{fig_tree3_schem}. The use of simulated data is to show the differences between the QP-Lasso and the proposed algorithm. In the subsequent section, data from a real network with the same structure as Fig. \ref{fig_tree3_schem} is used to illustrate the proposed algorithm. Before we present the results, some implementation details are given as follows:
\subsubsection*{Number of Unknowns} For the network in Fig. \ref{fig_tree3_schem}, $7$ unknowns are considered. This includes a leak and fault in each node except at node $1$, where the leak and fault are not distinguishable. Hence we consider only one unknown at this node (referred to as \textit{leak} in the following) for which the positivity constraints are not applied.

\subsection{Simulated data} 

 \begin{figure}
 \centering
 \includegraphics[trim={1cm 0cm 7.7cm 21.5cm},clip,width=0.99\linewidth]{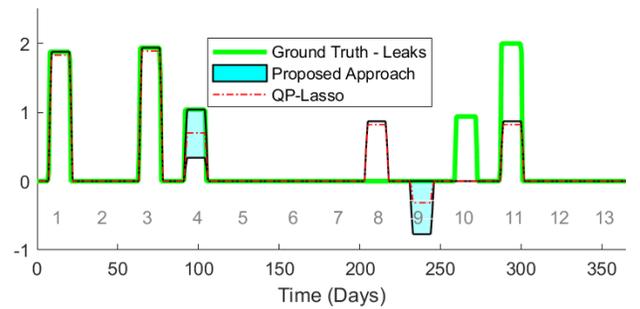}
 \caption{Leak Estimation results at Zone 1 (Simulated data)}
 \label{fig_sim_data_S1}
 \end{figure}

 \begin{figure}
 \centering
 \includegraphics[trim={1cm 0cm 7.7cm 17.5cm},clip,width=0.99\linewidth]{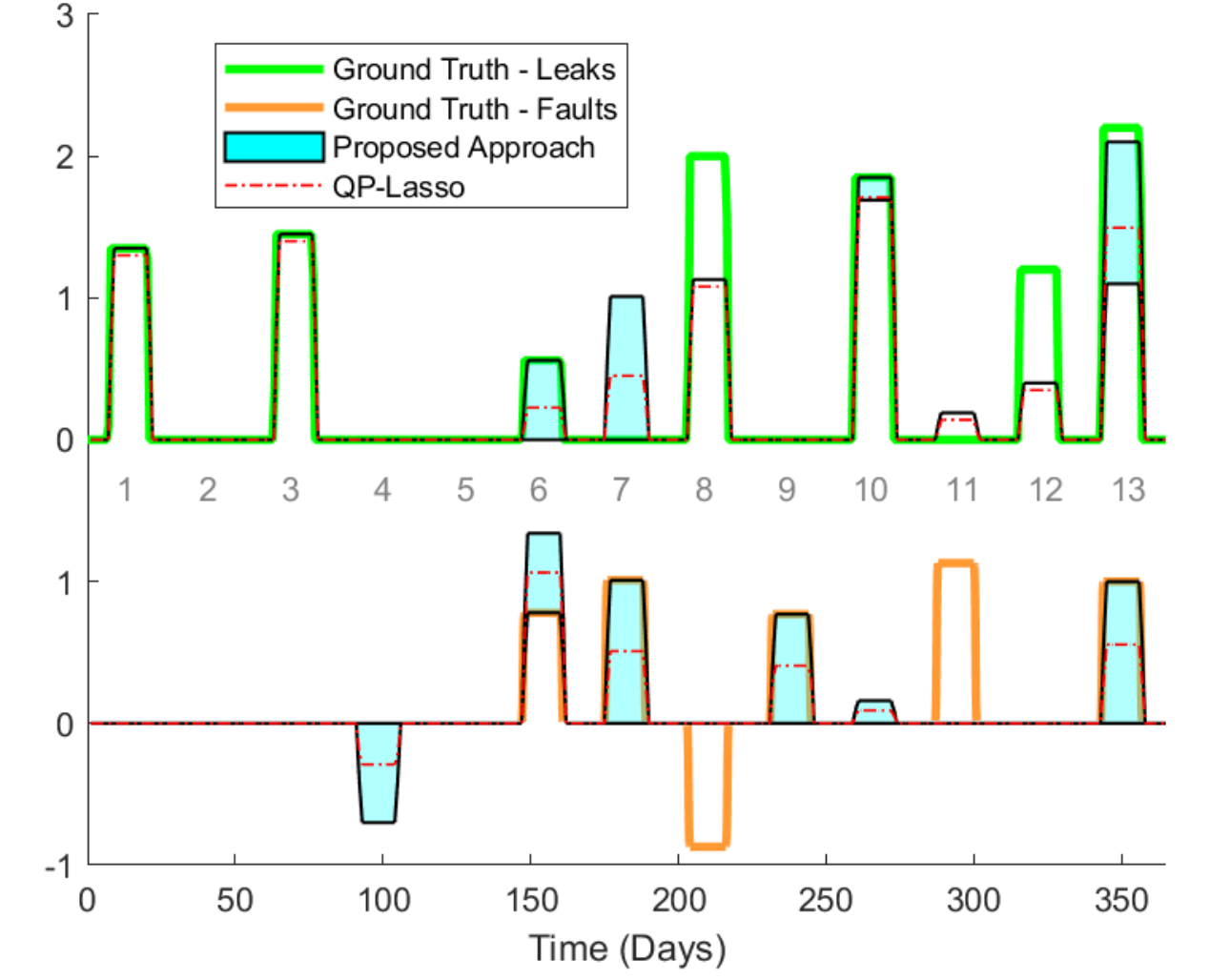}
 \caption{Leak/Fault Estimation at Zone 2 (Simulated data)}
 \label{fig_sim_data_S2}
 \end{figure}

 \begin{figure}
 \centering
 \includegraphics[trim={1cm 0cm 7.7cm 17.5cm},clip,width=0.99\linewidth]{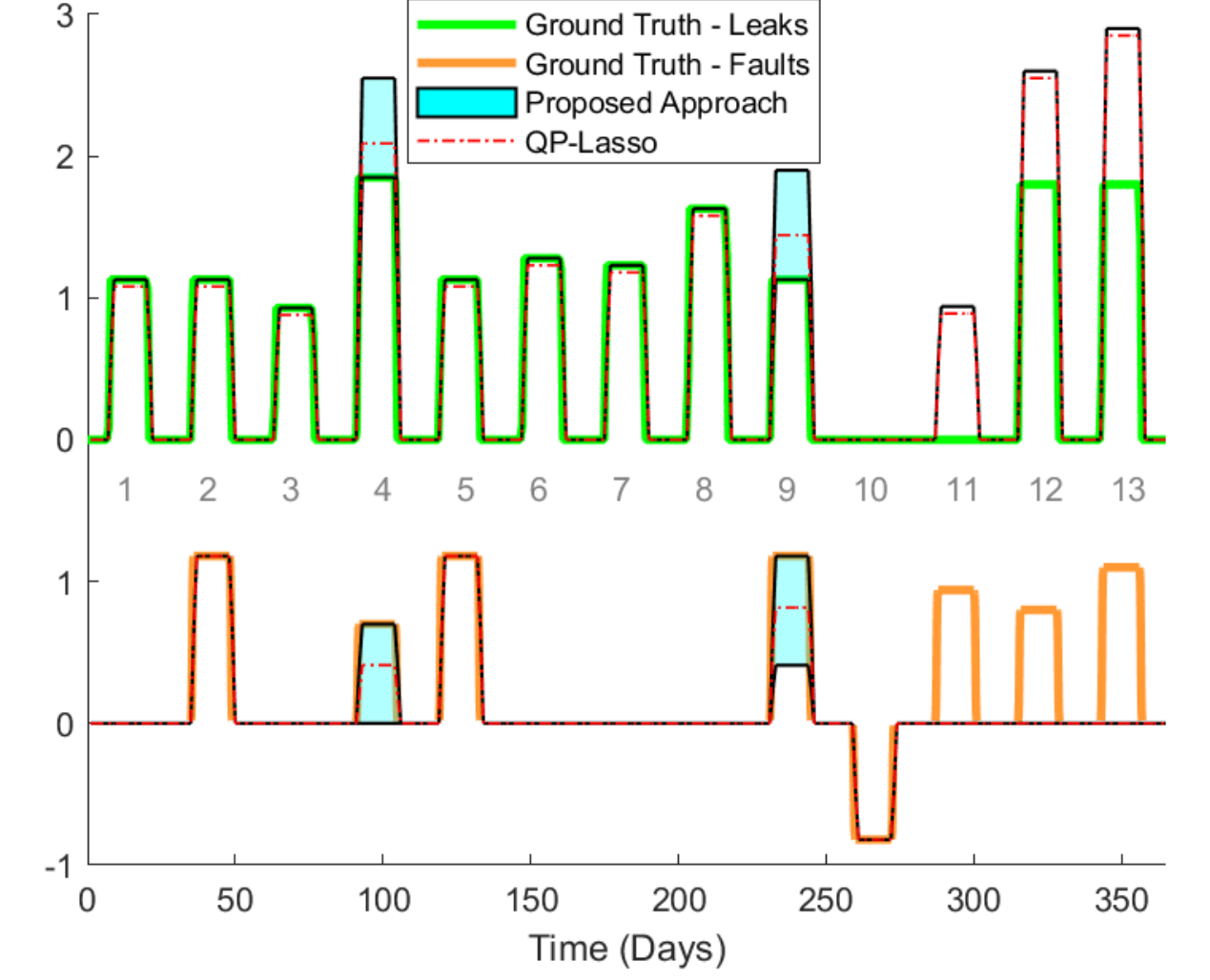}
 \caption{Leak/Fault Estimation at Zone 3 (Simulated data)}
 \label{fig_sim_data_S3}
 \end{figure}

 \begin{figure}
 \centering
 \includegraphics[trim={1cm 0cm 7.7cm 17.5cm},clip,width=0.99\linewidth]{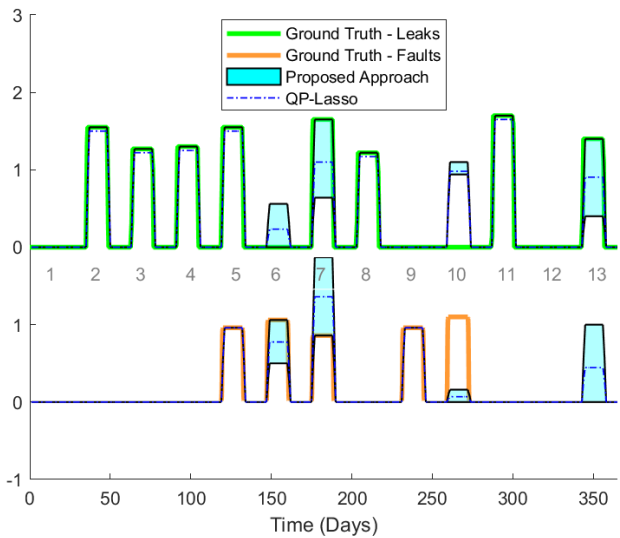}
 \caption{Leak/Fault Estimation at Zone 4 (Simulated data)}
 \label{fig_sim_data_S4}
 \end{figure}

We illustrate the differences between the QP-Lasso optimization and the proposed algorithm over data simulated from the topology given in Fig. \ref{fig_tree3_schem}. 

The fault-leak combinations of 4 elements is the maximum number of unknowns that can be estimated. We simulated all the $21$ detectable structures and $14$ undetectable ones. Further, we also simulated the $21$ undetectable structures of 5 faults. In order to provide a cleaner visualization, only a reduced number of simulations are displayed in Fig.~\ref{fig_sim_data_S1}-\ref{fig_sim_data_S4}. Note that, to show the leaks and faults in the same figure, we have a custom y-axis label that is set to zero at two y-axis points to indicate the reference for the two signals being shown. 

The estimation from the proposed algorithm is given by the black lines which sometimes encompass a shaded cyan region. The cyan region to illustrate the range over which the estimated leaks can take values that arises due to the $\ell_1$ norms of several fault-leak combinations being the same. 
\begin{itemize} \itemsep -1pt
	\item Cases $1$ to $7$, and $9$ represent the scenario where ${G_\mathcal{F}}$ is detectable and a solution of \eqref{est_problem}.
\item Cases $8$, $10$ and $11$ represent the scenario where ${G_\mathcal{F}}$ is detectable but is not a solution of  \eqref{est_problem} (other structures fit the data with smaller $\ell_1$ norm).
\item Case $12$ and $13$ represent the scenario where the true structure is undetectable.
\end{itemize}

In all cases, $\lambda=0.05$ was used as the parameter for the QP-Lasso implementation (using larger $\lambda$ values resulted larger bias in the results). Even though only $13$ combinations are displayed, the following conclusion remain true for all combinations of the simulated data.

\begin{itemize}
	\item If the solution to \eqref{est_problem} is unique, then QP-Lasso and Algorithm~\ref{alg_constraint} give identical results (up to numerical bias implied by the $\ell_1$ regularization).
	\item If \eqref{est_problem} has multiple solutions, then the unique QP-Lasso solution is always in the range of solutions provided by Algorithm~\ref{alg_constraint}.
	\item If the true structure ${G_\mathcal{F}}$ is a solution of \eqref{est_problem}, then it lies in the range of the solutions provided by Algorithm~\ref{alg_constraint}. This is represented by cases $1$ to $9$ in Fig.~\ref{fig_sim_data_S1}-\ref{fig_sim_data_S4}. 
	\item For each data point, QP-Lasso took $3.9 \text{ms}$ and the online part of the proposed algorithm took $0.09\text{ms}$. The one-time offline part of the algorithm to compute the detectable combinations took $16\text{ms}$. These simulation results were performed on the MATLAB computing environment using a PC running Intel i7-8550U processor.
\end{itemize}

This example shows that the proposed algorithm has at least the same performance over QP-Lasso at a much lower computational cost in the online process. At the same time, the algorithm provides the flexibility to understand the ambiguity in the final estimation providing a range of values over which the unknowns can lie.

\subsection{Real data} 
The data used in this paper comes from a French collaboration project, SPHEREAU\footnote{(in French) \url{https://www.hydreos.fr/projets/sphereau/44.html}}, which aims to optimize the overall functioning of a rural water distribution network. We study a rural collective-operated water distribution network in the GrandEst region in France. The network gets water from two sources with the pipes run about 300 km serving nearly 15000 habitants spread over 50 different communes. The instrumentation system is largely composed of water flow meters with level sensors augmenting them whenever there are water reservoirs. 

The dataset used in this paper is composed of 484 days of data with measurement every 15 minutes, and for the purposes of illustration and clarity, we estimate the faults every day over the averaged residue. Note that in practice, the prediction model $\mathcal{M}_i$ used in the generation of the graph $G_{\mathcal{FR}}$ are not \emph{a priori} available. In this study, these predicted data have been generated using a Reproducing Kernel Hilbert Space (RKHS) identified model to capture the characteristics of flow meters. More details on the forecasting model can be obtained from \cite{brentan2018infer}. Again, in case of zone 1, leaks and faults are not distinguishable and all anomalies are labelled leaks (which can take negative values). 

\subsubsection{Propagating estimated leaks} Before discussing the results of the algorithm, we describe another flexibility that arises from the use of graph representation. Consider the case when sensor data is missing or is stuck at constant value (that is, any case where a simple observation or data pre-processing can reveal fault). In these cases, it is possible to instruct the Algorithm 2 to further restrain the set of possible solutions as the ones containing the \emph{a priori} detected faults. 

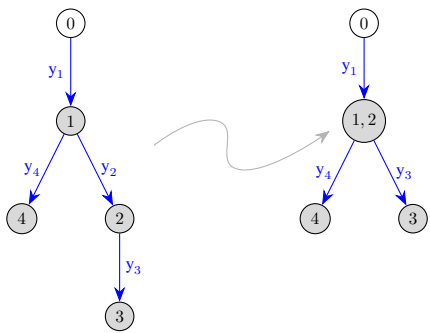
\begin{figure}[h]
\centering
 \begin{tikzpicture}[scale=0.65, every node/.style={scale=0.65}]
    \node at (0,0)[inner sep=3pt, draw, shape=circle] (nx){$0$};
    \node at (0,-2)[inner sep=3pt, draw, shape=circle, fill=gray!30] (n1) {$1$};
    \node at (1,-4)[inner sep=3pt, draw, shape=circle, fill=gray!30] (n2) {$2$};
    \node at (1,-6)[inner sep=3pt, draw, shape=circle, fill=gray!30] (n3) {$3$};
    \node at (-1,-4)[draw, shape=circle, fill=gray!30] (n4) {$4$};     
    \draw [->,blue] (nx) -- (n1) node[left, midway]{$\text{y}_1$};
    \draw [->,blue] (n1) -- (n2) node[right, midway]{$\text{y}_2$}; 
    \draw [->,blue] (n2) -- (n3) node[right, midway]{$\text{y}_3$}; 
    \draw [->,blue] (n1) -- (n4) node[left, midway]{$\text{y}_4$};    

	\draw [->,gray!60] plot[smooth, tension=1] coordinates {(1.7,-2.5) (3,-2) (3.5,-3) (5.3,-2.2)};		
    \node at (6,0)[inner sep=3pt, draw, shape=circle] (nx){$0$};
    \node at (6,-2)[inner sep=3pt, draw, shape=circle, fill=gray!30] (n1) {$1,2$};
    \node at (7,-4)[inner sep=3pt, draw, shape=circle, fill=gray!30] (n3) {$3$};
    \node at (5,-4)[draw, shape=circle, fill=gray!30] (n4) {$4$};     
    \draw [->,blue] (nx) -- (n1) node[left, midway]{$\text{y}_1$};
    \draw [->,blue] (n1) -- (n3) node[right, midway]{$\text{y}_3$}; 
    \draw [->,blue] (n1) -- (n4) node[left, midway]{$\text{y}_4$};        
  \end{tikzpicture}
\caption{Node merging due to missing data in Sensor 2}
\label{fig_topology_G_R}
\end{figure}

For example, take the case of the sensor measuring zone 2 consumption (or Sensor 2 for convenience) which has Sensor 1 upstream. If Sensor 2 is declared uninformative, then it is natural to remove it from the residual graph (just as if the sensor did not exist). Consequently, if leaks can be estimated for Sensor 1, they now represent leaks which are possibly physically located either in Zone 1 or in Zone 2. 
 In graph theoretic terms, this would be equivalent to the merging of the nodes $1 \text{ and } 2$ as shown in the Fig.~\ref{fig_topology_G_R}. This suggests that, in case of uninformative Sensor 2, leaks estimated at zone 1 provide an actual upper bound for possible leaks located either in zone 1 or in zone 2. We subsequently term this transaction as \emph{propagating estimated leaks} towards missing sensors.

\begin{figure*}
 \centering
 \includegraphics[trim={0cm 0cm 1.7cm 19.4cm},clip,scale=0.85]{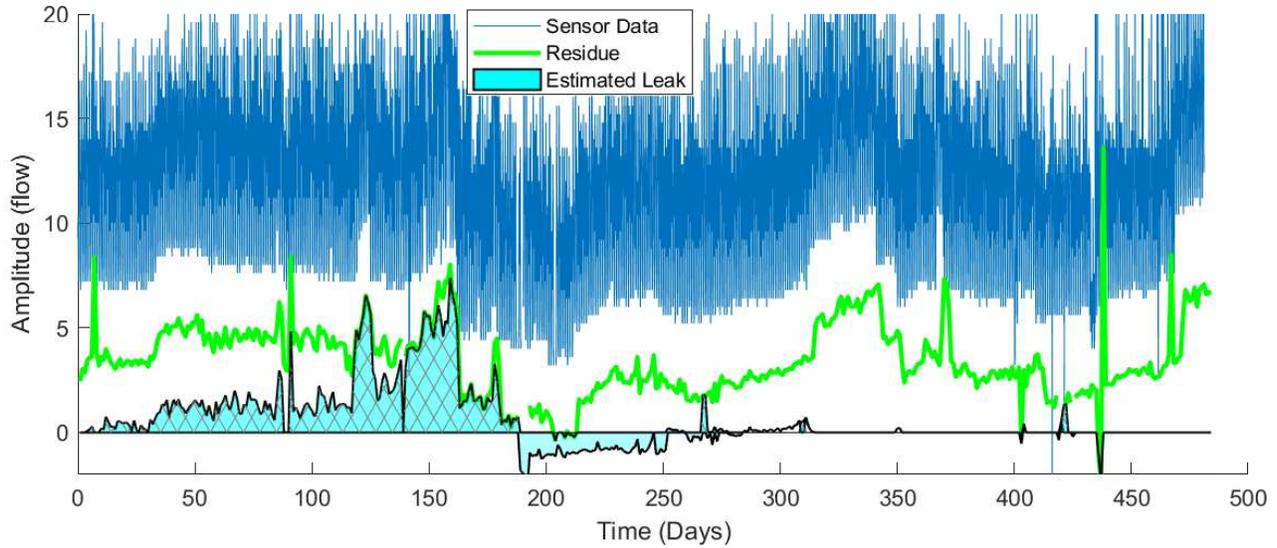}
 \caption{Leak Estimation at Zone 1 for the project data  (hatch : propagated leaks)}
 \label{fig_real_data_s1}
 \end{figure*}

\begin{figure*}
 \centering
 \includegraphics[trim={0cm 0cm 1.5cm 19.4cm},clip,scale=0.85]{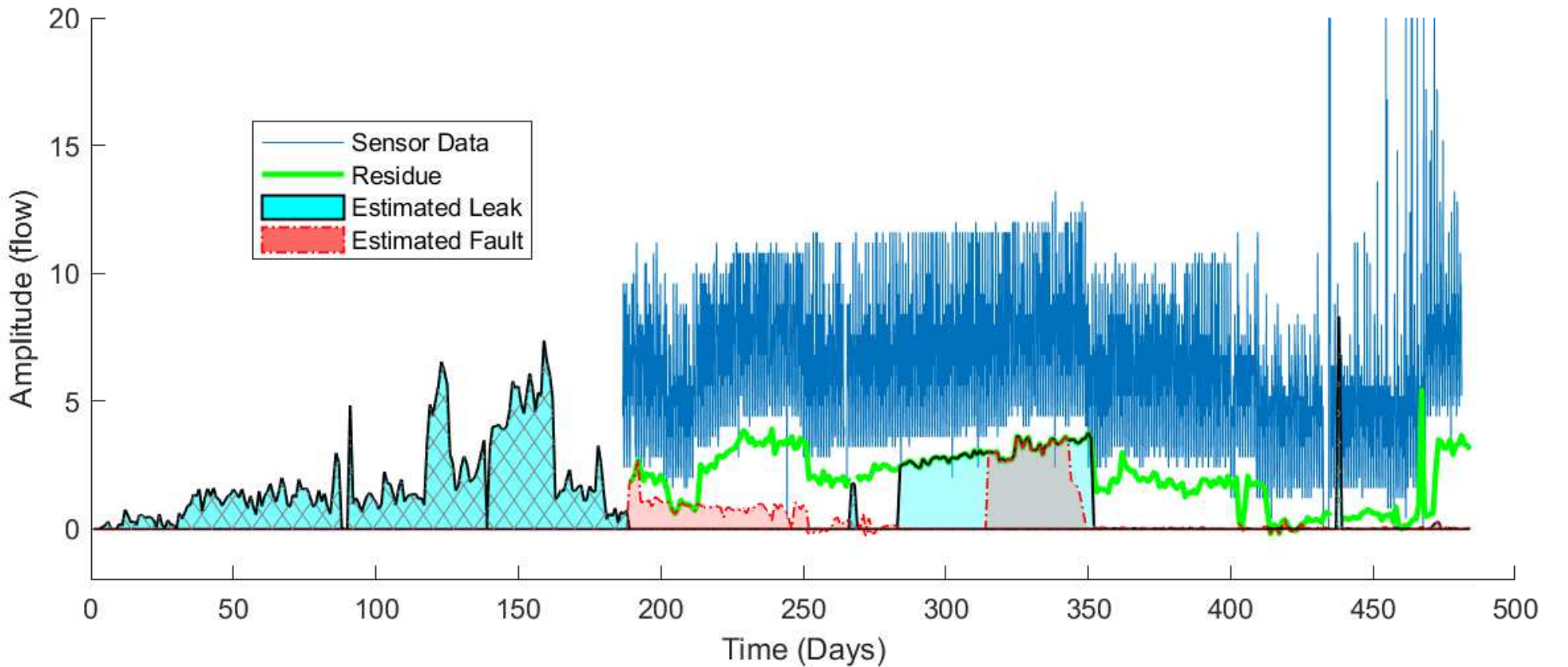}
 \caption{Leak/Fault Estimation at Zone 2 for the project data (hatch : propagated leaks)}
 \label{fig_real_data_s2}
 \end{figure*}

\begin{figure*}
 \centering
 \includegraphics[trim={0cm 0cm 1.7cm 19.4cm},clip,scale=0.85]{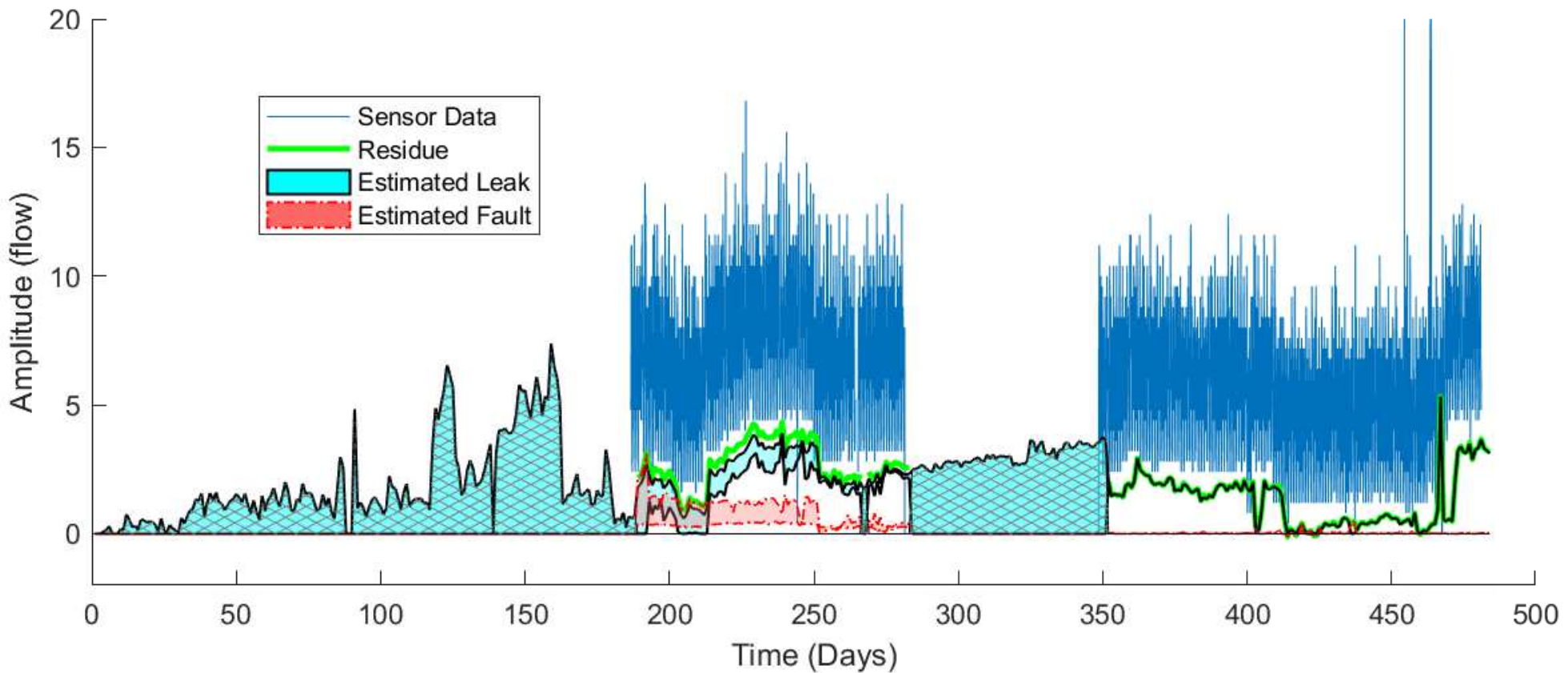}
 \caption{Leak/Fault Estimation at Zone 3 for the project data (hatch : propagated leaks))}
 \label{fig_real_data_s3}
 \end{figure*}

\begin{figure*}
 \centering
 \includegraphics[trim={0cm 0cm 1.7cm 19.4cm},clip,scale=0.85]{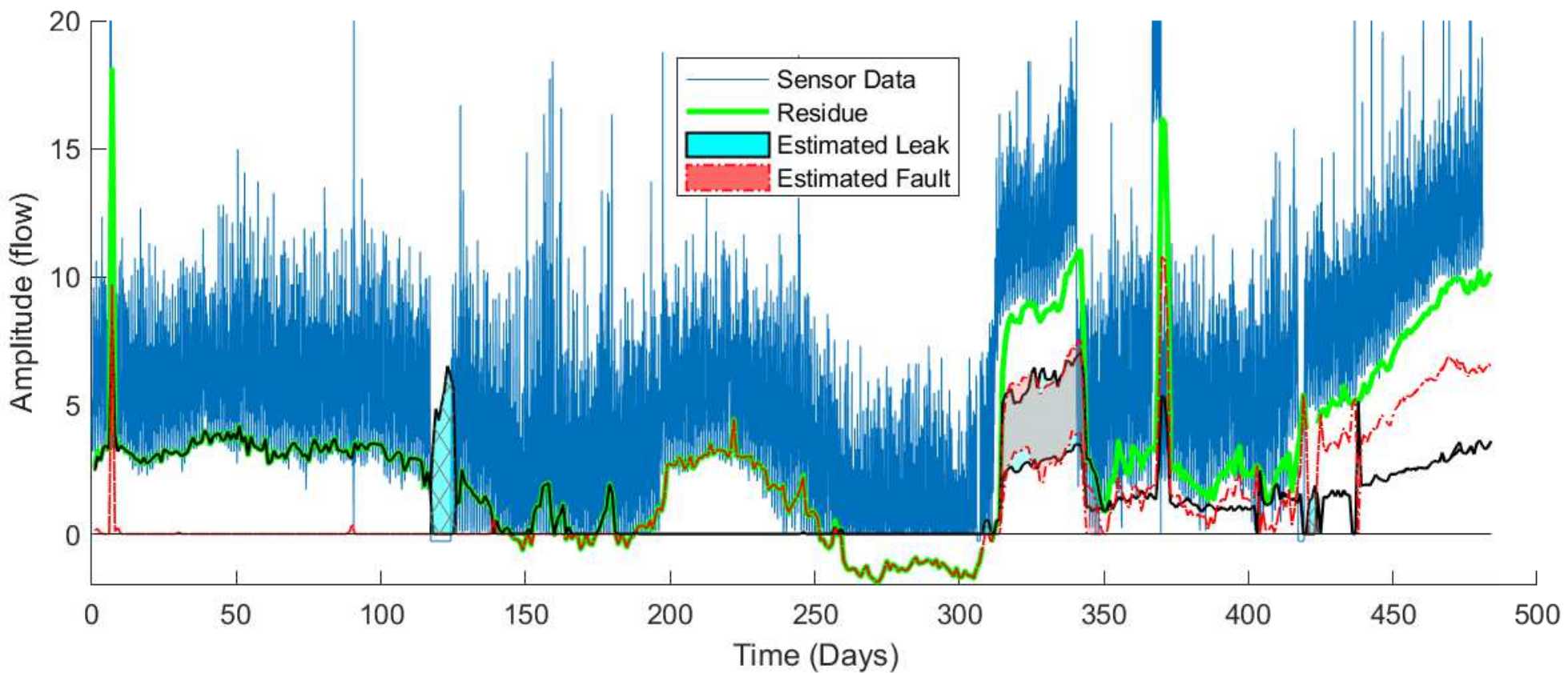}
 \caption{Leak/Fault Estimation at Zone 4 for the project data (hatch : propagated leaks)}
 \label{fig_real_data_s4}
 \end{figure*}

To illustrate the results of the algorithm, the sensor data, averaged residues obtained using the model predicted data, and the estimated leaks (and faults) at different zones are given in the Figs.~\ref{fig_real_data_s1}-\ref{fig_real_data_s4}. When data is uninformative (missing or stuck at 0), the estimated leaks are obtained through propagation from upstream sensors, as explained above, and is indicated using grey hatchings over the cyan shaded region. 

 
Though the real data did not have accurate ground truth, a few details are known. The faults on Sensor $1$, $2$, and $3$ are either missing data or stuck at a value. However, it is understood from the field engineers that the sensor measuring zone 4 experiences some drifts (to the point where it indicates more flow that is measured upstream).

We make the following observations from the results:
 
\subsubsection*{Handling missing data} 
Consider the region from  $\text{time}=0$  to $\text{time}=190$. During this period, data from Sensor 2 and 3 are missing (further, sensor 4 is missing around $\text{time}=120$). Here the graph is topologically modified enforcing faults in 2 and 3, and the leak estimated in  Sensor 1 are propagated to Sensors $2$ and $3$ (and 4 at $\text{time}=120$).

\subsubsection*{Detectable structures} Between $\text{time}=350$ and $\text{time}=484$, the situation represents the case where the fault structure is detectable and a solution of \eqref{est_problem}. The visualizations in the Figs.~\ref{fig_real_data_s1}-\ref{fig_real_data_s4} illustrate that the algorithm is able to disentangle  leaks and faults and properly assign them (according to field engineers). This can be used for triggering alerts or reconstruct the most plausible measure from combining the estimated leaks and the identified RKHS model.

\subsubsection*{Hybrid Scenario} In the region between $\text{time}=280$ and $\text{time}=350$ days, the Sensor 3 is down (stuck at zero). The true solution of the problem here is most probably (according to field engineers) that a leak is present in zone 3, a fault is present in Sensor 3, and at $\text{time}=320$ a leak appears in zone $4$ while Sensor $4$ also shows a drift. At this time, the true fault structure is detectable according to Theorem \ref{thm_detectability} but \eqref{est_problem} has several solutions and the true solution is not one of them. The solution of \eqref{est_problem} locates a range of leaks and faults spread out on sensor 2 and sensor 4. However, it can be seen that in this case, by means of \emph{a priori} fault detection on sensor 3 and using leak propagation, an improved solution is obtained that includes the true underlying solution. 


With these observations, we have shown that the Algorithm~\ref{alg_constraint} provides a flexible approach in estimating the leaks that can aid engineers by informing the uncertainty in the estimated leaks/faults and hence take more informed decisions. It is also possible to inform network managers that a given network is at its limit in terms of detectability and explicitly show which fault would imply undetectability, helping them to prioritize physical interventions on the network.

\section{Concluding Remarks} \label{sec_conclusions}
In this paper, we provided a novel modelling approach to capture process and sensor faults in systems modelled as a network using the graph representation. We derived graph-theoretic conditions for simultaneous detectability of these faults. We exploited these detectability results to develop a leak estimation algorithm based on a regularized least square problem, which shows more flexibility and lower computation than an equivalent quadratic programming implementation. The proposed approach has been tested both on simulated and real-data issued from a rural water distribution network.

Further work is intended both from fundamental and leak estimation application point of views. It would be imperative to extend the presented theory to cases of dynamical relationships between sensors or other types of networks (for example, when the initial graph is not a DAG due to the presence of loops or has multiple sources).  Further, the structural property of detectability can be used for the problem of placing sensors in the network to improve detectability for different scenarios, such as that in \cite{perelman2016sensor}. 

In terms of application to leak estimation, it will be useful to incorporate different types of sensors, for instance, pressure or level sensors in the water distribution network. Furthermore, the optimization criterion presented can be improved by incorporating other type of constraints which rely on temporal information. For example, consider that it is unlikely for several leaks or faults to appear at the exact same day and time. This leads to a constraint on how much consecutive estimations can differ. Finally, for large-scale networks, application of this algorithm faces a challenge: the flow amplitudes measured at the source and sink nodes will be considerably different posing practical challenges in using this approach to distinguish leaks and faults. Hence, a graph-partition approach that splits the global network into optimal subgraphs where the leak estimation approach can be directly applied would be of interest.

%
%

\ifCLASSOPTIONcaptionsoff
  \newpage
\fi



\bibliographystyle{IEEEtran}
\bibliography{IEEEabrv,./leakageRef}

\end{document}